# Large-area Epitaxial Monolayer MoS$_2$

Dumitru Dumcenco[1], Dmitry Ovchinnikov[1], Kolyo Marinov[1], Oriol Lopez-Sanchez[1], Daria Krasnozhon[1], Ming-Wei Chen[1], Philippe Gillet[2], Anna Fontcuberta i Morral[3], Aleksandra Radenovic[4], Andras Kis[1*]

[1]*Electrical Engineering Institute, Ecole Polytechnique Federale de Lausanne (EPFL), CH-1015 Lausanne, Switzerland*
[2]*Institute of Condensed Matter Physics, Ecole Polytechnique Federale de Lausanne (EPFL), CH-1015 Lausanne, Switzerland*
[3]*Institute of Materials, Ecole Polytechnique Federale de Lausanne (EPFL), CH-1015 Lausanne, Switzerland*
[4]*Institute of Bioengineering, Ecole Polytechnique Federale de Lausanne (EPFL), CH-1015 Lausanne, Switzerland*

*\*Correspondence should be addressed to: Andras Kis, andras.kis@epfl.ch*

**Two-dimensional semiconductors such as MoS$_2$ are an emerging material family with wide-ranging potential applications in electronics, optoelectronics and energy harvesting. Large-area growth methods are needed to open the way to the applications. While significant progress to this goal was made, control over lattice orientation during growth still remains a challenge. This is needed in order to minimize or even avoid the formation of grain boundaries which can be detrimental to electrical, optical and mechanical properties of MoS$_2$ and other 2D semiconductors. Here, we report on the uniform growth of high-quality centimeter-scale continuous monolayer MoS$_2$ with control over lattice orientation. Using transmission electron microscopy we show that the monolayer film is composed of coalescing single islands that share a predominant lattice orientation due to an epitaxial growth mechanism. Raman and photoluminescence spectra confirm the high quality of the grown material. Optical absorbance spectra acquired over large areas show new features in the high-energy part of the spectrum, indicating that MoS$_2$ could also be interesting for harvesting this region of the solar spectrum and fabrication of UV-sensitive photodetectors. Even though the interaction between the growth substrate and MoS$_2$ is strong enough to induce lattice alignment, we can easily transfer the grown material and fabricate field-effect transistors on SiO$_2$ substrates showing mobility superior to the exfoliated material.**

The most investigated member of transition metal dichalcogenides (TMDs) family, molybdenum disulfide (MoS$_2$) has attracted widespread attention for a variety of next generation electrical and optoelectronic device applications because of its unique properties. In the bulk, this material has a crystalline structure consisting of covalently bonded layers weakly coupled to each other by weak van der Waals forces. Owing to the weak coupling, two-dimensional (2D) monolayer MoS$_2$ can be easily obtained by exfoliation using scotch tape[1] or liquid-phase exfoliation[2,3]. Whereas bulk MoS$_2$ is a semiconductor with an indirect bandgap of 1.2 eV,[4] monolayer MoS$_2$ is a direct gap semiconductor with a bandgap of at least 1.8 eV[5-8] due to the 2D confinement.[8] It also has Raman-active modes that are very sensitive on the thickness which provides a convenient method for determining the number of layers with a reliable precision[9] in addition to atomic force microscopy and optical detection techniques.[10] The dramatic difference in the electronic structure of monolayer $MoS_2$ in comparison with its bulk counterpart offers many opportunities for the diverse applications. Using 2D crystals of $MoS_2$ mechanically exfoliated from bulk geological samples, versatile devices including field effect transistors with high on/off current ratio[10], memory cells[11], ultrasensitive photodetectors[12,13] and nanopores[14] have been reported. Because $MoS_2$ has high mechanical flexibility and breaking strength,[15] all these devices can in principle be implemented on flexible substrates.[16]

However, exfoliation of geological and samples grown using chemical vapor transport[17] lacks the systematic control of the thickness, size and uniformity of the 2D film and is not scalable for large-scale device fabrication. Because of this, several methods, such as decomposition of thiomolybdates[18], and sulfurization of Mo metal[19] or molybdenum oxide[20-22] have been exploited to synthesize $MoS_2$ on diverse substrates. Among them, chemical vapor deposition (CVD) is the most promising method to synthesize monolayer $MoS_2$ triangular islands tens micrometers in size[20-22]. In most of these reports, $SiO_2$ was used as the growth substrate, resulting in the random orientation of the $MoS_2$ domains because of the amorphous nature of the substrate and its relatively high surface roughness. This inevitably results in a large concentration of grain boundaries that can be detrimental to the electrical and mechanical properties of the grown films over length-scales exceeding several micrometers. In order to avoid this, it is necessary to control the crystallographic orientation of $MoS_2$ islands during growth so that they could coalesce into a uniform, single-crystal layer with a reduced number of grain boundaries.

Such control could in practice be achieved using a suitable atomically smooth crystalline substrate. In the case of classical three-dimensional materials this is normally achieved using epitaxial growth. Here, single-domain epitaxy can result in the deposition of a crystalline layer with a well-defined lattice orientation of the overlayer with respect to the growth substrate. This requires sufficiently strong interaction between the growth substrate and the deposited films. Because 2D materials are only weakly interacting via van der Waals bonds, this control is harder to implement. In one hand, this allows the growth of layered materials on top of each other without the usual requirement for lattice matching using van der Waals epitaxy[23] but it also makes it more difficult to control the lattice orientation of the deposited films. This was indeed the case in previously reported deposition of centimeter-scale monolayer $MoS_2$ films that have been deposited on other layered materials such as nearly lattice-matched mica[24-27] or graphene[28] where $MoS_2$ grains showed random orientation with respect to each other and the substrate lattice.

Here, we use highly polished, EPI-ready grade sapphire substrates commonly used as growth substrates for GaN in LED manufacturing to achieve control over lattice orientation during CVD growth of monolayer $MoS_2$. Due to the commensurability of the sapphire lattice with $MoS_2$, single-crystal domains show crystallographic alignment. This is the first time control over lattice orientation with centimeter-scale uniformity has been achieved during the growth of a 2D semiconductor and it opens the way to large-area growth of high-quality single-crystal $MoS_2$.

We start by preparing the c-plane sapphire surface by annealing it in air for 1h at a temperature of 1000 °C just prior to the growth process.[29] A detailed AFM-based analysis shows that the sapphire surface is characterized by atomically flat terraces ~50-70 nm wide with a step height of 0.22 nm due to a small miscut angle estimated to be ~0.2°. The terrace edges follow the [11$\bar{2}$0] direction on the average[30]. Annealed samples are transferred to the CVD system and growth is performed based on the gas-phase reaction between $MoO_3$ and sulphur evaporated from solid sources[21,22] using ultrahigh-purity argon as the carrier gas.

The growth procedure results in characteristic single-crystal domains in the shape of well-defined equilateral triangles that merge into a continuous monolayer film covering a typical area of 6 mm x 1 cm in the middle portion of the growth substrate, Figure 1a. Figure 1b presents optical images of regions showing partial (top image) and almost full (bottom) coverage. A careful examination of a region with incomplete coverage, figure 1c, reveals that most of the single-crystal domains are oriented along dominant directions. A reflection high-energy electron diffraction (RHEED) pattern is shown in the inset of figure 1c. The appearance of streaks typical of 2D materials and the $MoS_2$ (1×1) pattern indicate the growth of a homogeneous and well-structured film over a large area. The vast majority of single-crystal domains are well aligned with the relative orientation of edges that can be expressed as multiples of 60°. This is confirmed by the orientation histogram presented on figure 1f for the same area as on figure 1c, showing that the dominant edge orientations are 0° and ±60°. A small fraction of domains show edges with a relative angle of 90°.

We find that the sharp edges of monolayer $MoS_2$ single crystals form angles in the 8-11° range with respect to terrace edges, figure 1d. According to previous STM studies, these $MoS_2$ edges are expected to be (10$\bar{1}$0) Mo zig-zag edges terminated with sulphur[31]. The sapphire terrace step height of 0.22 nm is sufficiently low to assure the growth of continuous $MoS_2$ single crystals as confirmed by TEM imaging and electrical transport measurements. Figure 1e shows a high-resolution TEM image of a freestanding membrane[32] of monolayer $MoS_2$ showing its regular atomic structure with a clearly discernible hexagonal symmetry. White spots correspond to holes in the atomic structure while dark spots correspond to sulphur and molybdenum atoms.

We proceed to analyze the large-scale crystal structure and their relative orientation using bright-field TEM imaging and selected-area electron diffraction. Figure 2a shows a low-magnification image of group of aligned $MoS_2$ triangular islands. We closely examine a small region involving two neighboring islands by positioning the select-area aperture at the red circle containing two islands with their edges forming a 60° angle and acquiring the diffraction pattern shown in the inset of Figure 2a. Even though this diffraction pattern has been acquired from both triangles, we can only observe one set of diffraction spots with six-fold symmetry, showing the precise alignment of lattices in both of the islands. We have repeated the same analysis on more than 10 similar cases with the same result. Figure 2b shows another example of a less frequently occurring situation with two islands merging under a 90° angle. In this case we can clearly distinguish two sets of [$\bar{1}$100] diffraction spots.

The asymmetry of the Mo and S sublattices also allows us to determine the orientation of the $MoS_2$ lattice with respect to the island edges[22]. Figure 2c schematically illustrates two possible relative orientations for monolayer $MoS_2$ growing epitaxially on the atomically smooth sapphire. The driving force behind these two orientations is presumably the dipole-dipole interaction between the sulphur and aluminum atoms. On the left side, we have the predominant orientation in which the $MoS_2$ island and the sapphire lattice are commensurate. On the right-hand side, we show the much less likely configuration that could lead to the appearance of neighboring triangular islands with edges forming a 90° angle.

We checked the quality of our large-area monolayer MoS$_2$ films by performing optical characterization. Figure 3a shows typical Raman spectra of the CVD material compared to monolayer MoS$_2$ exfoliated onto sapphire. The spectra show two characteristic peaks in this spectral range: the out-of-plane vibration of S atoms ($A'_1$) with a Raman shift of ~405 cm$^{-1}$ and the in plane vibration of Mo and S atoms ($E'$) at ~385 cm$^{-1}$.[9] The observed frequency difference confirms that the deposited material is monolayer MoS$_2$. The ratio between the $A'_1$ and $E'$ mode intensities which can be used as an indicator of doping levels[33] indicates that the CVD material is less doped than the exfoliated counterpart while the $A'_1$ mode is narrower in the CVD material. This points to the CVD material having a higher degree of order and less doping, presumable due to a reduced concentration of sulphur impurities.

Figure 3b shows a typical photoluminescence spectrum acquired at room temperature on CVD-grown and adhesive tape exfoliated monolayer MoS$_2$. We can clearly resolve the intense A excitonic peak at 659 nm (1.88 eV). Typical peak widths are ~ 26 nm (~72 meV) and smaller than in exfoliated MoS$_2$ samples (~40 nm or ~111 eV), indicating that our CVD MoS$_2$ has superior optical qualities to the exfoliated material. Relative PL intensities also indicate a reduced doping level in the CVD material. Detailed photoluminescence mapping of single domains did not resolve any internal structure indicating the absence of internal grain boundaries.

Optical transparency of the sapphire substrate and the large area of the sample covered by monolayer MoS$_2$ allow us to perform UV-visible absorption characterization using a simple bench-top spectrophotometer. The resulting spectrum shown on Figure 3c represents broadband absorbance measurements on MoS$_2$. The spectrum shows the well-known A and B excitonic absorption bands at 695 nm (1.78 eV) and 611 nm (2.02 eV)[6,7]. This demonstrates the high optical quality and uniformity of our monolayer MoS$_2$ over a large area. In addition, we can also clearly observe the recently reported C peak at 430 nm (2.88 eV)[34] as well as the D peak at 303 nm (4.08 eV). Whereas the A and B peaks are associated to optical absorption by band-edge excitons, peaks C and D are associated with van Hove singularities[35] of monolayer MoS$_2$. The fact that MoS$_2$ shows enhanced absorbance in the 200-500 nm range could play an important role in designing solar cells and UV-sensitive photodetectors.

We further characterize our CVD material by performing electrical transport measurements. We first transfer the CVD-grown material from the sapphire substrate to a degenerately doped silicon substrate acting as a back-gate with a thermally grown 270 nm thick layer of SiO$_2$. Field-effect transistors based on single domains are transferred using a combination of electron-beam lithography followed by the deposition of 90-nm-thick gold electrodes. Selected devices were etched into the shape of 6-terminal Hall-bar structures (Figure 4a) by oxygen plasma. The devices were then annealed at 200°C in Ar atmosphere to eliminate resist residues and reduce contact resistance. In a second annealing step the devices were annealed in vacuum (5×10$^{-7}$ mbar) for ~15 hours at 130-140°C in order to remove water and other adsorbates from the surface of the 2D semiconducting channel[27]. Electrical measurements were performed immediately after this in vacuum. Figure 4b shows the current vs. bias voltage ($I_{ds}$ vs. $V_{ds}$) characteristics of our device for several different values of the gate voltage $V_g$. The observed linear $I_{ds}$-$V_{ds}$ characteristics indicates the high quality of contacts and the absence of significant charge injection barriers at room temperature. Figure 4c shows the transfer characteristics ($I_{ds}$ vs $V_g$) of the device from Figure 4a recorded for a bias voltage $V_{ds} = 2$ V. The source-drain distance is $L = 6.7$ μm, $l_{12} = 4$ μm is the distance between the voltage probes $V_1$ and $V_2$ and the device width $W = 8.3$ μm. From this curve, we can obtain the field-effect mobility $\mu_{FE}$ defined as $\mu_{FE} = \left[ dG / dV_{bg} \right] \times \left[ l_{12} / (WC_{ox}) \right]$ where $G$ is the four-probe conductance defined as $G = I_{ds}/(V_1-V_2)$ with $I_{ds}$ the drain current, $V_1 - V_2$ the measured voltage difference between the voltage probes and $C_{ox} = 1.3 \times 10^{-4}$ F/m$^2$ the capacitance between the channel and the back gate per unit area. At high gate voltages we observe the mobility reach a value of 43 cm²/Vs, comparable to results from previous two-terminal measurements on CVD MoS$_2$ (ref. 36) and superior to devices based on exfoliated MoS$_2$ [25,27].

In conclusion, we have achieved epitaxial chemical vapor deposition growth of monolayer MoS$_2$ with high degree of control over lattice orientation by using atomically smooth surfaces of sapphire as the growth substrate. The large-area, centimeter-scale MoS$_2$ is formed from merging single-crystalline domains with the majority of them having the same lattice orientation, resulting in long-range crystalline order. A low fraction of domains shows a 90° relative orientation. The high quality of our sample is further demonstrated by optical and electrical measurements showing properties superior to exfoliated samples. The high degree of uniformity of our film also allowed optical absorbance measurements to be performed in a broad wavelength range, showing the presence of high-energy absorbance peaks that indicate the suitability of MoS$_2$ for harvesting the green and blue regions of the solar spectrum and for the fabrication of photodetectors operating in this wavelength range. The proposed growth strategy involving the use of atomically smooth sapphire substrates could pave the way for large area growth of MoS$_2$ with high optical and electrical quality retained over large length scales, allowing its use in future electronic and optoelectronic devices.

**METHODS**

**Growth procedure**

Monolayer MoS$_2$ has been grown by chemical vapor deposition (CVD) on sapphire c-plane. After consecutive cleaning by acetone/isopropanol/DI-water the substrates were annealed for 1h at 1000°C. After that, they were placed face-down above a crucible containing ~5 mg MoO$_3$ (≥99.998% Alfa Aesar) and loaded into a 2-inch split-tube three-zone CVD furnace. CVD growth[20] was performed at atmospheric pressure using ultrahigh-purity argon as the carrier gas. A second crucible containing 350 mg of sulfur (≥99.99% purity, Sigma Aldrich) was located upstream from the growth substrates. The growth recipe is following: set the temperature of 300°C with 200 sccm for 10 min, ramp to 700°C with a rate of 50°C min$^{-1}$ and 10 sccm of carrier gas flow, set the temperature to 700°C for 10 min, cool down to to 570°C with 10 sccm gas flow, increase the gas flow to 200 sccm and open the furnace for rapid cooling.

**TEM and AFM imaging**

CVD MoS$_2$ was transferred from sapphire using the wet transfer KOH method. Samples were first spin coated at 1500 rpm with PMMA A2, resulting in a ~100nm thick polymer film. These were detached in a 30% KOH solution, washed several times in DI water and transferred onto TEM grids. TEM grids were annealed in the flow of Ar and $H_2$ for 8 hours at 400 °C in order to remove the polymer film. For low-resolution imaging and diffraction studies, 10 nm thick $Si_3N_4$ windows were used while for high resolution TEM (HR-TEM) we used PELCO Holey Silicon Nitride Support Film with 2.5 um holes in a 200 nm – thick $Si_3N_4$ support. Transmission electron microscopy was performed using JEOL 2200 FS operated in the 120 - 200 keV energy range. High-resolution TEM (HR-TEM) images were recorded at a magnification of 1M×. A series of 10-30 images was recorded and stacked with drift correction and averaging using the Stackreg plugin in Fiji (ImageJ). Island orientations in optical images were analysed using the directionality plugin in Fiji (ImageJ). Samples were also imaged using an atomic force microscope (Asylum Research Cypher) operating in AC mode.

**Electrical characterization**

CVD-grown single domains of $MoS_2$ were transferred using PMMA A2 as a support film and etching in 30% KOH onto degenerately doped Si substrate covered with 270 nm $SiO_2$. The PMMA film is dissolved in acetone and residues are removed by annealing in Ar atmosphere at 350°C for 5 hours. PMMA A4 was used as the etching mask during oxygen plasma etching. The devices were then annealed at 200°C in Ar atmosphere to eliminate resist residues and reduce contact resistance.

**ACKNOWLEDGEMENTS**

Thanks go to D. Walters and J. Cleland from Asylum Research for suggestions related to sapphire surface preparation. We thank R. Gaal for technical assistance with the Raman setup, T. Heine for useful discussions, M. G. Friedl for technical assistance and A. Allain for help with electrical transport measurements. Device fabrication was carried out in the EPFL Center for Micro/Nanotechnology (CMI). We thank Z. Benes (CMI) for technical support with e-beam lithography and D. Alexander (CIME) for support with electron microscopy. This work was financially supported by Swiss SNF Sinergia Grant no. 147607, funding from the European Union's Seventh Framework Programme FP7/2007-2013 under Grant Agreement No. 318804 (SNM), Marie Curie ITN network "MoWSeS" (grant no. 317451), ERC grant no. 240076 and SNF Grant 135046.

# REFERENCES

1. Novoselov, K. S. *et al.* Two-dimensional atomic crystals. *PNAS* **102**, 10451-10453, doi:10.1073/pnas.0502848102 (2005).
2. Coleman, J. N. *et al.* Two-Dimensional Nanosheets Produced by Liquid Exfoliation of Layered Materials. *Science* **331**, 568-571, doi:10.1126/science.1194975 (2011).
3. Smith, R. J. *et al.* Large-Scale Exfoliation of Inorganic Layered Compounds in Aqueous Surfactant Solutions. *Adv. Mater.* **23**, 3944-3948, doi:10.1002/adma.201102584 (2011).
4. Kam, K. K. & Parkinson, B. A. Detailed photocurrent spectroscopy of the semiconducting group VIB transition metal dichalcogenides. *J. Phys. Chem.* **86**, 463-467, doi:10.1021/j100393a010 (1982).
5. Lebegue, S. & Eriksson, O. Electronic structure of two-dimensional crystals from ab initio theory. *Phys. Rev. B* **79**, 115409 (2009).
6. Splendiani, A. *et al.* Emerging Photoluminescence in Monolayer $MoS_2$. *Nano Lett.* **10**, 1271-1275, doi:10.1021/nl903868w (2010).
7. Mak, K. F., Lee, C., Hone, J., Shan, J. & Heinz, T. F. Atomically Thin $MoS_2$: A New Direct-Gap Semiconductor. *Phys. Rev. Lett.* **105**, 136805, doi:10.1103/PhysRevLett.105.136805 (2010).
8. Kuc, A., Zibouche, N. & Heine, T. Influence of quantum confinement on the electronic structure of the transition metal sulfide $TS_2$. *Phys. Rev. B* **83**, 245213, doi:10.1103/PhysRevB.83.245213 (2011).
9. Lee, C. *et al.* Anomalous Lattice Vibrations of Single- and Few-Layer $MoS_2$. *ACS Nano* **4**, 2695-2700, doi:10.1021/nn1003937 (2010).
10. Benameur, M. M. *et al.* Visibility of dichalcogenide nanolayers. *Nanotechnology* **22**, 125706, doi:10.1088/0957-4484/22/12/125706 (2011).
11. Bertolazzi, S., Krasnozhon, D. & Kis, A. Nonvolatile Memory Cells Based on MoS2/Graphene Heterostructures. *ACS Nano* **7**, 3246-3252, doi:10.1021/nn3059136 (2013).
12. Lopez-Sanchez, O., Lembke, D., Kayci, M., Radenovic, A. & Kis, A. Ultrasensitive photodetectors based on monolayer $MoS_2$. *Nat. Nanotechnol.* **8**, 497-501, doi:10.1038/NNANO.2013.100 (2013).
13. Yin, Z. *et al.* Single-Layer $MoS_2$ Phototransistors. *ACS Nano* **6**, 74-80, doi:10.1021/nn2024557 (2012).
14. Liu, K., Feng, J., Kis, A. & Radenovic, A. Atomically Thin Molybdenum Disulfide Nanopores with High Sensitivity for DNA Translocation. *ACS Nano*, doi:10.1021/nn406102h (2014).
15. Bertolazzi, S., Brivio, J. & Kis, A. Stretching and Breaking of Ultrathin $MoS_2$. *ACS Nano* **5**, 9703-9709, doi:10.1021/nn203879f (2011).
16. Chang, H.-Y. *et al.* High-Performance, Highly Bendable $MoS_2$ Transistors with High-K Dielectrics for Flexible Low-Power Systems. *ACS Nano*, doi:10.1021/nn401429w (2013).
17. Schäfer, H. *Chemical transport reactions*. (Academic Press, 1964).
18. Liu, K.-K. *et al.* Growth of Large-Area and Highly Crystalline $MoS_2$ Thin Layers on Insulating Substrates. *Nano Lett.* **12**, 1538-1544, doi:10.1021/nl2043612 (2012).
19. Zhan, Y., Liu, Z., Najmaei, S., Ajayan, P. M. & Lou, J. Large-Area Vapor-Phase Growth and Characterization of $MoS_2$ Atomic Layers on a $SiO_2$ Substrate. *Small* **8**, 966-971, doi:10.1002/smll.201102654 (2012).

20  Lee, Y.-H. *et al.* Synthesis of Large-Area MoS$_2$ Atomic Layers with Chemical Vapor Deposition. *Adv. Mater.* **24**, 2320-2325, doi:10.1002/adma.201104798 (2012).
21  Najmaei, S. *et al.* Vapour phase growth and grain boundary structure of molybdenum disulphide atomic layers. *Nat. Mater.* **12**, 754-759, doi:10.1038/nmat3673 (2013).
22  van der Zande, A. M. *et al.* Grains and grain boundaries in highly crystalline monolayer molybdenum disulphide. *Nat. Mater.* **12**, 554-561, doi:10.1038/nmat3633 (2013).
23  Koma, A. Van der Waals epitaxy for highly lattice-mismatched systems. *J. Cryst. Growth* **201**, 236-241 (1999).
24  Ji, Q. *et al.* Epitaxial Monolayer MoS$_2$ on Mica with Novel Photoluminescence. *Nano Lett.* **13**, 3870-3877, doi:10.1021/nl401938t (2013).
25  Radisavljevic, B. & Kis, A. Mobility engineering and a metal-insulator transition in monolayer MoS2. *Nat. Mater.* **12**, 815-820, doi:10.1038/NMAT3687 (2013).
26  Jariwala, D. *et al.* Band-like transport in high mobility unencapsulated single-layer MoS$_2$ transistors. *Appl. Phys. Lett.* **102**, 173107, doi:10.1063/1.4803920 (2013).
27  Baugher, B., Churchill, H. O. H., Yang, Y. & Jarillo-Herrero, P. Intrinsic Electronic Transport Properties of High Quality Monolayer and Bilayer MoS2. *Nano Lett.*, doi:10.1021/nl401916s (2013).
28  Shi, Y. *et al.* van der Waals Epitaxy of MoS$_2$ Layers Using Graphene As Growth Templates. *Nano Lett.* **12**, 2784-2791, doi:10.1021/nl204562j (2012).
29  Yoshimoto, M. *et al.* Atomic-scale formation of ultrasmooth surfaces on sapphire substrates for high-quality thin-film fabrication. *Appl. Phys. Lett.* **67**, 2615-2617, doi:doi:http://dx.doi.org/10.1063/1.114313 (1995).
30  Curiotto, S. & Chatain, D. Surface morphology and composition of c-, a- and m-sapphire surfaces in O2 and H2 environments. *Surface Science* **603**, 2688-2697, doi:http://dx.doi.org/10.1016/j.susc.2009.07.004 (2009).
31  Lauritsen, J. V. *et al.* Size-dependent structure of MoS2 nanocrystals. *Nat Nano* **2**, 53-58 (2007).
32  Brivio, J., Alexander, D. T. L. & Kis, A. Ripples and Layers in Ultrathin MoS$_2$ Membranes. *Nano Lett.* **11**, 5148-5153, doi:10.1021/nl2022288 (2011).
33  Chakraborty, B. *et al.* Symmetry-dependent phonon renormalization in monolayer MoS$_2$ transistor. *Phys. Rev. B* **85**, 161403 (2012).
34  Klots, A. R. *et al.* Probing excitonic states in ultraclean suspended two-dimensional semiconductors by photocurrent spectroscopy. *arXiv preprint*, 1403.6455 (2014).
35  Britnell, L. *et al.* Strong Light-Matter Interactions in Heterostructures of Atomically Thin Films. *Science* **340**, 1311-1314, doi:10.1126/science.1235547 (2013).
36  Schmidt, H. *et al.* Transport Properties of Monolayer MoS$_2$ Grown by Chemical Vapor Deposition. *Nano Lett.*, doi:10.1021/nl4046922 (2014).

**FIGURES**

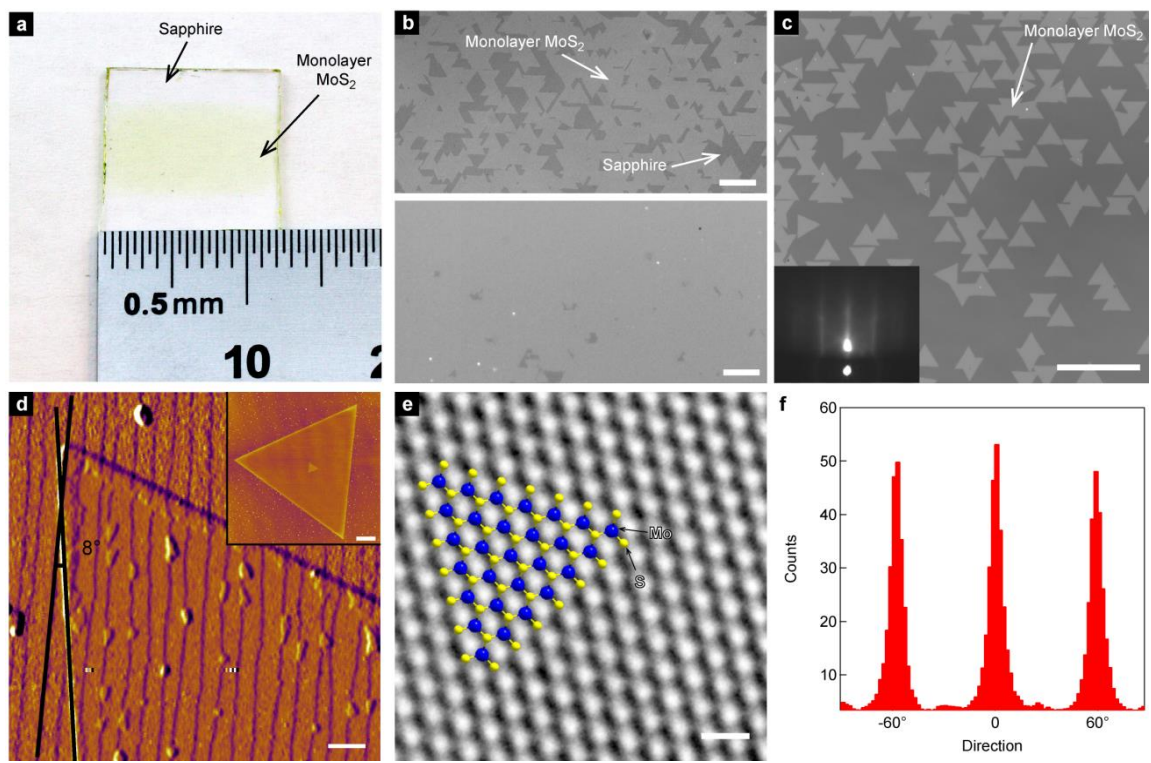

**Figure 1. Monolayer MoS$_2$ growth with controlled lattice orientation. a,** Photograph of cm-scale monolayer MoS$_2$ grown on sapphire **b,** Optical microscopy images from different regions of the sample showing incomplete coverage close to the edges and an almost complete coverage close to the center of the growth substrate. Scale bar length is 20 µm on the top image and 10 µm for the bottom image. Original optical images were converted to greyscale and the contrast was enhanced **c,** Optical microscopy image of monolayer MoS$_2$ grains grown on atomically smooth sapphire. Scale bar length is 50 µm. Inset: RHEED pattern acquired on the CVD-grown sample showing a film with long-range structural order. **d,** Atomic force microscope image showing the growth of a monolayer MoS$_2$ island across step edges due to the miscut of the sapphire surface. Scale bar length is 50 µm. Inset: AFM image of a monolayer MoS$_2$ grain showing sharp edges. Scale bar is 2 µm long. **e,** High-resolution TEM image of a suspended MoS$_2$ film showing the crystallinity of the sample. A top view of the structural model is overlaid. Scale bar is 0.5 nm long. **f,** Orientation histogram based on the area shown in part c confirms that the majority of MoS$_2$ grain edges are oriented along 0 and ±60° angles.
9

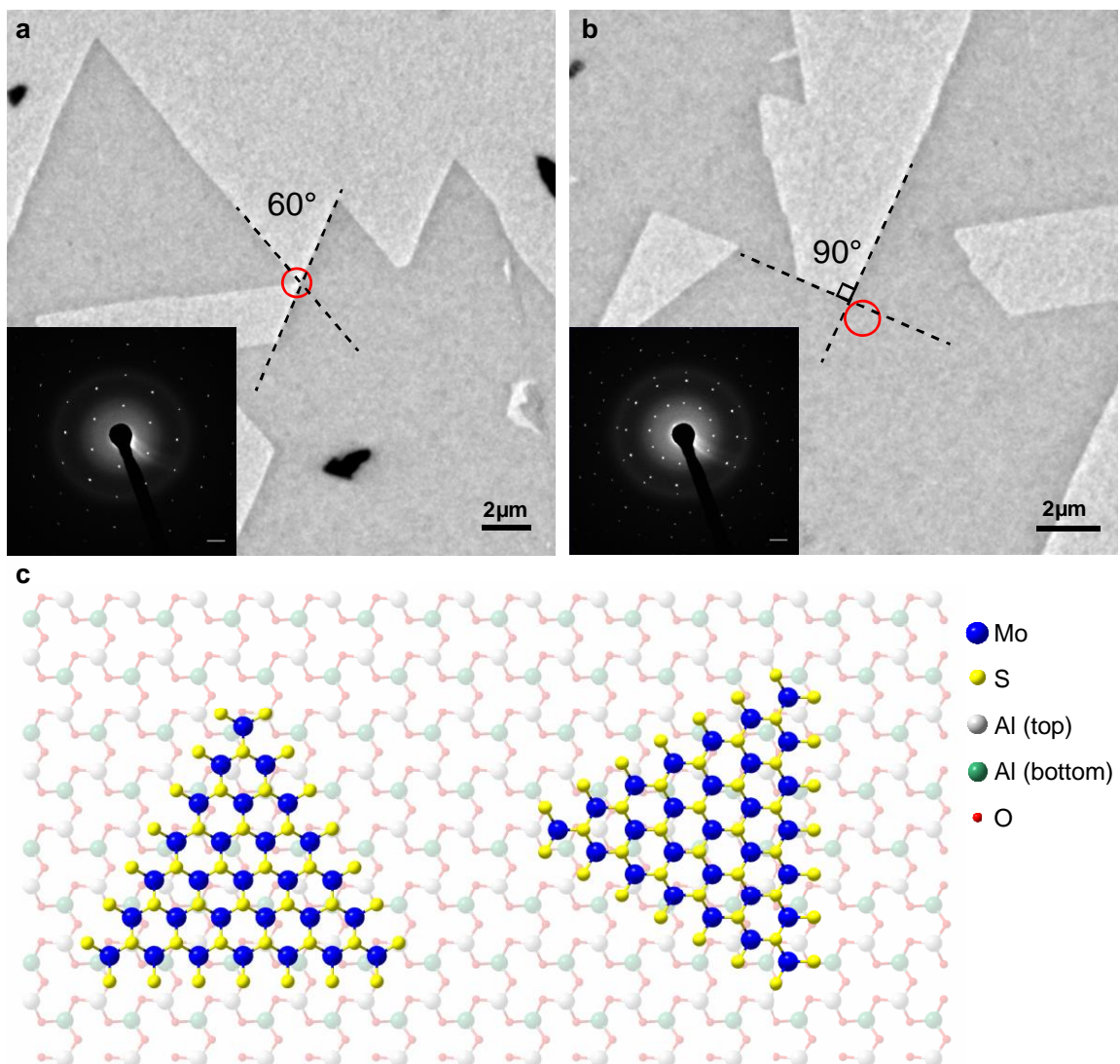

**Figure 2. Diffraction patterns from different island orientations. a,** Low-magnification TEM image of several neighboring MoS$_2$ islands. The diffraction pattern acquired from the area denoted with the red circle is shown in the inset and corresponds to the most common situation where the edges of with neighboring islands are oriented at 60° angles. Only one set of diffraction spots can be detected from such islands, indicating that their crystalline lattices are aligned. **b,** Low-magnification TEM image and the corresponding diffraction patterns from two merging islands with their edges forming a 90° degree angle. Two sets of diffraction spots, indicating a 30° relative lattice orientation can be seen here. **c,** A schematic drawing showing the top-view of relative lattice orientations between monolayer MoS$_2$ and c-plane sapphire. In the case of the arrangement of the left-hand side, the two lattices are commensurate. The MoS$_2$ island on the right-hand side has been rotated by a 30° angle, resulting in a 90° angle between the edges of neighboring islands

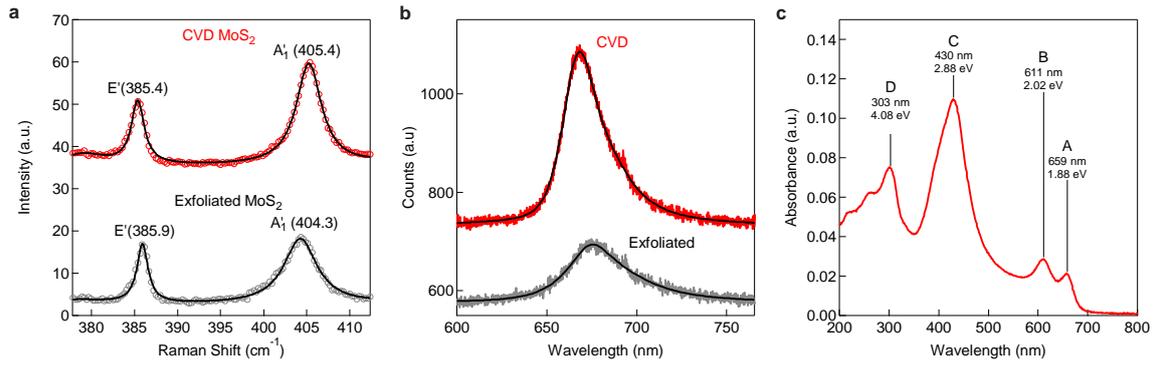

**Figure 3. Optical properties of large-area monolayer MoS$_2$. a,** Raman spectra of as-grown monolayer MoS$_2$ on sapphire and monolayer MoS$_2$ exfoliated from bulk crystals and transferred onto sapphire. **b**, Photoluminescence spectra of as-grown monolayer MoS$_2$ on sapphire and monolayer MoS$_2$ exfoliated from bulk crystals onto sapphire. Black lines in a and b correspond to fits. **c,** UV-Vis optical absorbance spectra acquired from large area monolayer MoS$_2$ showing the A and B absorption peaks due to band-edge excitons as well as C and D peaks associated with van Hove singularities of MoS$_2$.

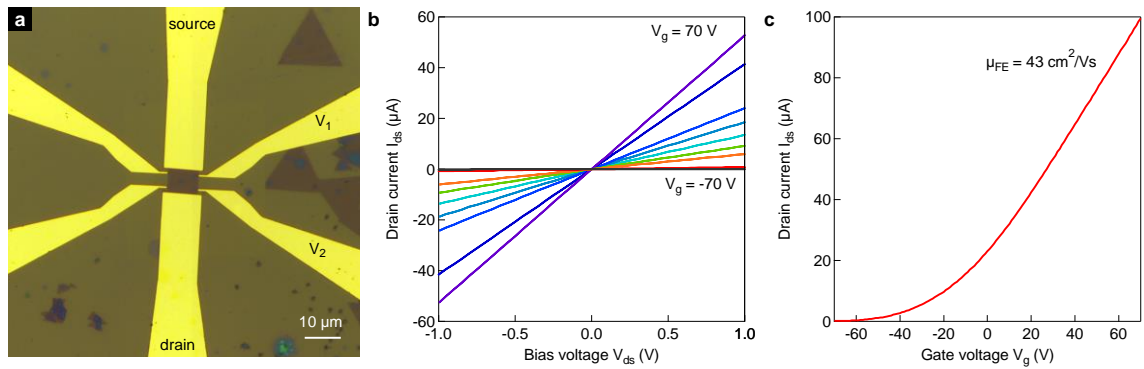

**Figure 4. Electrical properties of large-area monolayer MoS$_2$. a,** Optical image of a Hall-bar device based on a single-crystal island of MoS$_2$ transferred onto SiO$_2$. **b,** Sweeps of current $I_{ds}$ *vs.* bias voltage $V_{ds}$ characteristics of the device shown in a, indicating ohmic-like behavior of the contacts. **c,** Current as a function of gate voltage shows n-type behavior. A maximum field-effect mobility $\mu_{FE}$ = 43 cm$^2$/Vs can be extracted.